\documentclass[aps,prl,twocolumn,showpacs]{revtex4-1}

\usepackage{amsmath}
\usepackage{amssymb}
\usepackage{graphics}
\usepackage{epsfig}
\usepackage{multirow}
\usepackage{color}

\begin{document}

\title{Coexistence of bulk superconductivity and charge density wave in Cu$_x$ZrTe$_{3}$}

\author{Xiangde Zhu}
\altaffiliation{present address: High Magnetic Field Laboratory, Chinese
Academy of Sciences, Hefei 230031, P. R. China}
\author{Hechang Lei}
\author{C. Petrovic}
\affiliation{Condensed Matter Physics and Materials Science
Department, Brookhaven National Laboratory, Upton, New York 11973,
USA}

\date{\today}

\begin{abstract}
We report coexistence of bulk superconductivity and charge density wave (CDW) with superconducting critical temperature T$_c$ = 3.8 K in Cu intercalated quasi-two-dimensional crystals of ZrTe$_{3}$. The Cu intercalation results in the expansion of the unit cell orthogonal to the Zr-Zr metal chains ($\widehat{b}$ - axis) and partial filling of CDW energy bandgap without obvious shift of CDW transition temperature. $\widehat{b}$ - axis resistivity $\rho_b$ is not related to CDW, and its dominant scattering mechanism for both ZrTe$_{3}$ and Cu$_{0.05}$ZrTe$_{3}$ is the electron - electron Umklapp scattering.
\end{abstract}


\maketitle
Peierls instability results in the periodic modulation of electron density in
solids and generates an energy bandgap at the Fermi surface (FS), which leads to insulating behavior in one-dimensional (1D) metal\cite{1}. It can be observed as a charge density wave (CDW) superstructure in many low dimensional materials\cite{2}. Superconductivity (SC), another instability of FS, will also generate a bandgap at FS, which results in a zero resistance. There is a view that SC and CDW orders can only compete with each other. NbSe$_3$, whose structure is composed of infinite NbSe$_6$ trigonal prismatic chains along the monoclinic $\widehat{b}$ - axis, indures two CDW transition with CDW transition temperature (T$_{CDW}$) = 145 K (T$_1$) and 59 K (T$_2$). For example, High pressure continuously suppresses the second CDW transition and induces SC in quasi 1D NbSe$_3$\cite{3}; The suppression of charge order in Ba$_{1-x}K$$_x$BiO$_3$ via doping brings out SC with highest T$_c$ $\sim$ 30 K\cite{4}; Recently, both Cu intercalation\cite{5} and high pressure \cite{6} in TiSe$_{2}$ continuously suppress the CDW order, and induce SC with a dome-like phase diagram, which strongly supports the competing view. However, CDW and SC can coexist with each other, like in 2H-NbSe$_2$ with T$_{CDW}$ = 33 K and T$_c$ = 7.2 K\cite{2}. Angle-resolved photoemission spectroscopy (ARPES) results of 2H-NbSe$_2$ show that maximized SC at points in momentum space are directly connected by the CDW ordering vector, demonstrating that charge order can boost SC in electron-phonon coupled system\cite{7}, which is in direct contrast to the previous view.

The interplay between CDW and SC is reminiscent of SC in heavy fermion and the cuprate oxide phase diagram when the magnetic order is tuned by pressure ($P$) or doping to $T\rightarrow 0$\cite{8,9,10}. Superconducting mechanism in such electronic systems is likely to be mediated by the magnetic fluctuations\cite{11,12}. Similarly, the dome-like structure of T$_{c}(x)$ and the pairing mechanism in Cu$_{x}$TiSe$_{2}$ is argued to stem from the type of quantum criticality related to fluctuations in CDW order\cite{13}. Quantum phase transition is proposed to drive CDW order into a quantum nematic phase\cite{14,15}. On the other hand, the positive slope of the dome could be explained via the shift of the Fermi level caused by Cu intercalation and the negative slope by the enhanced
scattering for the high Cu concentration\cite{16}. This is also supported by
the evidence for a single gap s-wave order parameter, implying conventional
mechanism despite the dome-like evolution of T$_{c}$($x$)\cite{17}.
However, a dome-like structure of T$_{c}(P)$ was also discovered using $P$
as a tuning parameter and suggesting that impurity effects may not be
responsible for the dome closure\cite{6}.

Chalcogenide superconductors represent a weak coupling side of the
smectic order, akin to stripe order in cuprates\cite{14}. There is a
mounting evidence that in such systems CDW states transform into Fermi
liquid through an intermediate phase\cite{18}. Therefore it is of interest
to study the melting of CDW order parameter and possible nematic phases in a
superconductor with highly related and tunable two dimensional electronic
system.

ZrTe$_3$, as an trichalcogenide (MX$_3$, where M is the IV-VI transition-metal, and X is S, Se, Te), accommodates the quasi 2D ZrSe$_3$ type structure (shown in Fig. 1 (a))\cite{19,20}, rather than the quasi 1D NbSe$_3$ type structure. The crystallographic $\widehat{b}$ - axis defines the direction of two sets of ZrTe$_6$ trigonal prismatic chains with the shortest Zr-Zr distance. ZrTe$_3$ endures a CDW transition at T$_{CDW}$ $\sim $ 63 K with a CDW wave vector $\overrightarrow{q}$ $\equiv (\frac{1}{14},0,\frac{1}{3})$\cite{21,22}, and show corresponding hump anomaly on $\widehat{a}$ - axis resistivity ($\rho_a$) and $\widehat{c}$ - axis resistivity ($\rho_c$) around T$_{CDW}$\cite{21}. The FS of ZrTe$_{3}$ contains one quasi 1D electron like sheet (pocket) of states of $5p$ Te chain origin that is
responsible for CDW distortion and pseudogap formation at T $\gg $ T$_{CDW}$\cite{20,23,24,25}. ZrTe$_{3}$ exhibits filamentary SC below 2 K with no diamagnetic response in H = 10 Oe\cite{26}. The CDW and SC in ZrTe$_3$ exhibit rich and so called "competing" interplay under high pressure. With increasing $P$, the T$_{CDW}$ initially increases below 2 GPa and then decreases up to the vanishing point T$_{CDW}$ $\sim $ 40 K at $\sim$5 GPa\cite{27}; the T$_c$ of filamentary SC decreases and vanish at $P\sim$ 0.5 GPa, then a reentrant SC emerges at $\sim$5 GPa with T$_{c}$ rising monotonously up to 11 GPa (T$_{c}$ $\sim$ 4.5 K at 11 GPa). Keeping in mind that the FS of NbSe$_3$ mainly originates from the Nb 3d electrons and dome-like structure of T$_c(P)$ in TiSe$_2$, ZrTe$_3$ provides a different opportunity to study the interplay between CDW and SC. Although intercalation attempts have been done in alkali atoms intercalated into ZrSe$_{3}$\cite{28} and Cu intercalated ZrTe$_3$ via electrochemical method\cite{29}, no SC was discovered.

In this letter, we report coexistence of bulk SC and CDW in Cu intercalated ZrTe$_{3}$. The Cu intercalation results in the expansion of the unit cell orthogonal to $\widehat{b}$ - axis and partial filling of CDW bandgap without obvious shift of T$_{CDW}$. Anisotropic parameters of the superconducting state are presented. $\widehat{b}$ - axis resistivity ($\rho_b$) in the normal state is not related to CDW, and its dominant scattering mechanism for both ZrTe$_{3}$ and Cu$_{0.05}$ZrTe$_{3}$ is the electron - electron Umklapp scattering.

\begin{figure}
\includegraphics[width=0.45\textwidth,angle=0]{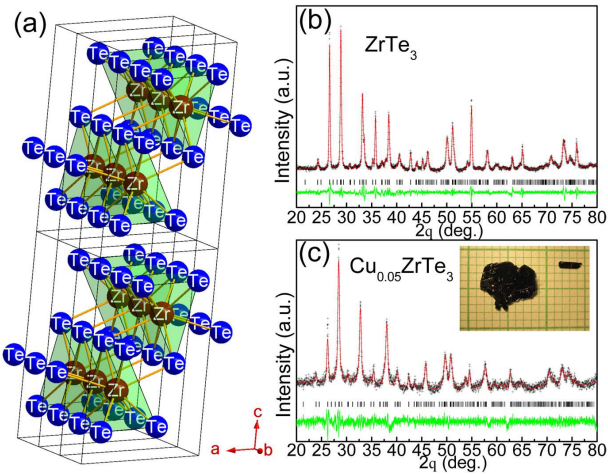}
\caption{
(a)The crystal structure of ZrTe$_{3}$ with blue (brown) symbols for Te (Zr) atoms. Comparison of the observed
(crosses) and calculated (solid line) powder X ray patterns of ZrTe$_{3}$
(b) and Cu$_{0.05}$ZrTe$_{3}$ (c). The bars \ (black) correspond to the Bragg
reflections and the lowest solid line is the difference between the observed
and the calculated patterns. }
\end{figure}

Single crystals of Cu$_{0.05}$ZrTe$_{3}$ and ZrTe$_{3}$ with typical size of 1%
$\times $3$\times $0.06 mm$^{3}$ (inset of Fig. 1(c)) elongated along the $\widehat{b}$ - axis were
grown via iodine vapor transport method from pure elements sealed in an evacuated quartz tube.
After 2 days pre-reacting at 973 K, the furnace gradient was kept between 1023 K and 923 K. Elemental
analysis was performed by using energy-dispersive X-ray spectroscopy in an
JEOL JSM-6500 scanning electron microscope. The Cu content is determined as the average of the different points on several crystals. Resistivity and magnetization were measured in Quantum Design PPMS-9 and MPMS-XL-5, respectively.
The observed (Cu $K\alpha $ radiation of Rigaku Miniflex) and calculated\cite{30} (Rietica software) powder X ray diffraction (XRD) patterns for ZrTe$_{3}$ and Cu$_{0.05}$ZrTe$_{3}$ are shown in Fig. 1 (b) and (c),
respectively. Both materials can be indexed to the ZrTe$_{3}$, indicating that the Cu intercalation does not change the crystal structure. The unit cell refinement yields lattice parameters a = 0.586(3) nm, b = 0.392(7) nm, c = 1.009(5) nm and $\beta $= 97.75(1)$^{\circ }$ for ZrTe$_{3}$, whereas for Cu$_{0.05}$ZrTe$_{3}
$ we obtained a = 0.588(2) nm, b = 0.392(8) nm, c = 1.011(1) nm, and $\beta $= 97.75(1)$^{\circ }$.
The Cu intercalation leads to slight expansion of the $\widehat{a}$ and $\widehat{c}$ - axis parameters.
It should be noted that Zr - Zr distances along the $\widehat{b}$ - axis chains are unchanged\cite{20}.

\begin{figure}
\includegraphics[width=0.45\textwidth,angle=0]{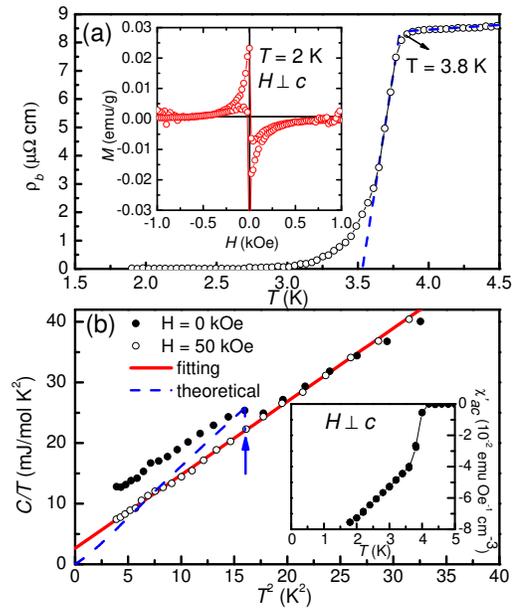}
\caption{(a) $\protect\rho _{b}$ (T) of
Cu$_{0.05}$ZrTe$_{3}$ near T$_{c}$. Inset shows the M-H loop measured at T = 2 K for H$\perp$c. (b) (C/T) of Cu$_{0.05}$ZrTe$_{3}$ as a function of T$^{2}$ for H = 0 (solid circles) and H$\parallel$c = 50 kOe (open circles). Inset shows temperature dependence of $\protect\chi_{ac}^{\prime }$. See text for details.}
\end{figure}

The onset of SC at T = 3.8 K can be observed in $\rho_b$, and the magnetization measurements confirm the SC in Cu$_{0.05}$ZrTe$_3$ (Fig. 2(a)). The shape of the magnetic hysteresis loop of is typical type - 2 inhomogeneous superconductors with some electromagnetic granularity (inset of Fig. 2(a)). This is in agreement with the transition from the temperature dependence of the real part of ac magnetic susceptibility ($\chi _{ac}^{\prime }$) of Cu$_{0.05}$ZrTe$_{3}$ (inset of Fig. 2(b)), and is similar to data in some high T$_{c}$ cuprates and SmFeAsO$_{0.85}$F$_{0.15}$\cite{31,32}. Although the $\chi _{ac}^{\prime }$ does not saturate down to 1.8 K, the large estimated superconducting volume fraction at T = 1.8 K (4$\pi \chi_{ac}^{\prime }$ $\sim$ 94\%) reveals the bulk nature of SC in Cu$_{0.05}$ZrTe$_{3}$, as opposed to the pure material\cite{26}.

In order to obain the electron-phonon coupling constant $\lambda$ , specific heat (C) of Cu$_{0.05}$ZrTe$_3$ was measured. Figure 2(b) depicts the specific heat divided by T (C/T) as a function of T$^{2}$ of Cu$_{0.05}$ZrTe$_{3}$ in H = 0 kOe and H$\parallel$c = 50 kOe. Superconducting transition is suppressed in 50 kOe and from its linear fit (solid line depicted in Fig. 2 (b)) C/T = $\gamma$ + $\beta $T$^{2}$ ($\gamma$T is electron contribution, and $\beta $T$^{3}$ is phonon contribution), we obtain $\gamma $= 2.64 $\pm$ 0.12 mJ/ molK$^{2}$
and $\beta$ = 1.21 $\pm$ 0.01 mJ/molK$^{4}$. A Debye temperature $\Theta _{D}$ = 186(1) K
can be estimated using $\Theta _{D} = [\frac{1.944\times 10^{6}\times N}{\beta
}]^{1/3}$ where N is the number of atoms per formula unit. From the McMillan
formula\cite{33}
\begin{equation}
\lambda =\frac{\mu ^{\ast }\ln (\frac{1.45T_{C}}{\Theta _{D}})-1.04}{%
1.04+\ln (\frac{1.45T_{C}}{\Theta _{D}})(1-0.62\mu ^{\ast }),}
\end{equation}
we estimate the $\lambda$ $\sim $0.68 by
assuming $\mu ^{\ast }$= 0.13 for the Coulomb pseudopotential. This is a
typical value of an intermediate coupling BCS superconductor. A specific
heat jump ($\Delta $C$/\gamma $T$_{c}$) can be observed around T = 4 K (marked by an arrow) confirming bulk SC.
The dashed curve depicted in Fig.2(b) is the calculated result of the isotropic BCS gap with 2$\Delta /$k$_{B}
$T$_{c}$ = 3.53 and $\Delta $C$/\gamma $T$_{c}$=1.43. Although the C data in the superconducting state is not good due to the inhomogeneity, a $\Delta $C$/\gamma $T$_{c}$ larger than 1.43 can be estimated, suggesting the intermediate or strong $e-p$ coupling.

\begin{figure}
\includegraphics[width=0.45\textwidth,angle=0]{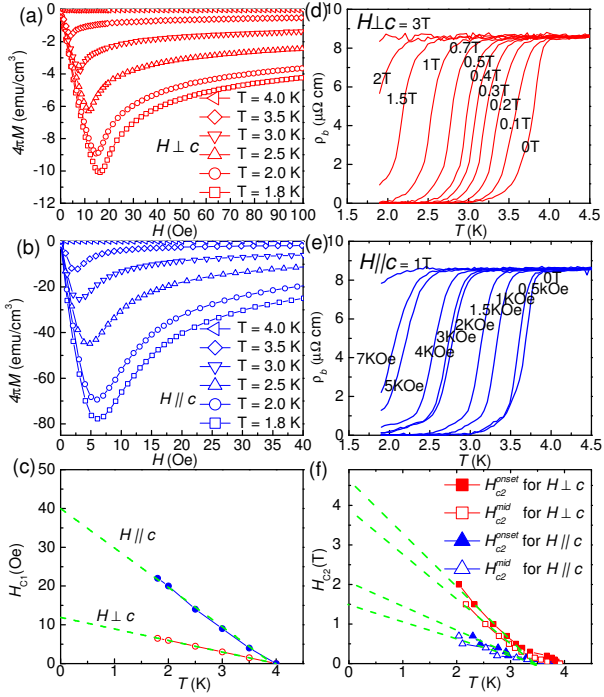}
\caption{Field dependence of magnetization of Cu$_{0.05}$ZrTe$_{3}$ for H$\bot$c (a) and H$\Vert$c (b). (c)
Temperature dependence of H$_{c1}$ for H$\bot$c and H$\Vert $c. The
dashed lines are the fitted lines. $\protect\rho _{b}$(T) of Cu$_{0.05}$%
ZrTe$_{3}$ for H$\bot$c (d) and H$\Vert$c (e). (f) H$_{c2}$(T) onset and H$_{c2}$(T) midpoint (see
text) of Cu$_{0.05}$ZrTe$_{3}$.}
\end{figure}

The anisotropic properties of Cu$_{0.05}$ZrTe$_3$ in the superconducting state are investigated. Figures 3(a) and (b) show field dependence of magnetization of Cu$_{0.05}$ZrTe$_{3}$
measured at different temperatures for H$\perp$c and H$\parallel$c. For H$\parallel$c,
the H$_{c1}$ is obtained using the demagnetization correction\cite{34}.
The H$_{c1}$(T) shows almost linear realtion rather than H$_{c1}$= H$_{c1}$(0)[1-(T/T$_{c}$)$^{2}$]
temperature dependence for both field orientation (Fig. 3(c), dashed lines).
Thus, we estimate H$_{c1}$(0) = $-0.693\frac{\partial Hc1}{\partial T}T|_{T=T_{c}}$
from the Werthamer-Helfand-Hohenberg (WHH) equation\cite{35} and we obtain
8.3(1) Oe (H$\perp$c) and 27.7(4) Oe (H$\parallel$c). Figures 3(d) and (e) show $\rho _{b}$(T) of
Cu$_{0.05}$ZrTe$_{3}$ measured in different fields for H$\perp $c and H$\parallel $c. Figure 3(f) plots
the temperature dependence of H$_{c2}^{onset}$ (onset) and H$_{c2}^{mid}$
(midpoint, 50\% of $\rho $ in the normal state). H$_{c2}$(0) are estimated
from the linear $H_{c2}$-T relation with WHH equation.
Due to small variance in the in plane resistivity values\cite{21}, we assume that coherence lengths
are isotropic in the $\widehat{a}$ - $\widehat{b}$ plane ($\xi _{a} \sim \xi
_{b}=\xi _{ab}$) and we estimate\cite{36} coherence leghts using $%
H_{c2}^{c}(0)=\Phi _{0}/(2\pi \xi _{ab}^{2})$ and $H_{c2}^{ab}(0)=\Phi
_{0}/(2\pi \xi _{ab}\xi _{c})$ where $\Phi _{0}$ is the flux quantum. The
Ginzburg - Landau (GL) parameters $\kappa _{i}$ and penetration depths are
estimated from $H_{c2}^{i}(0)/H_{c1}^{i}(0)=2\kappa _{i}^{2}/\ln \kappa _{i}$
(i = ab, c) and $\kappa_{ab}(0)=\lambda _{ab}(0)/\xi _{c}(0)$, respectively.
Finally, from $H_{c}(0)=H_{c1}^{ab}(0)/\sqrt{2}\kappa _{ab}(0)$, we estimate the
thermodynamic critical field. The anisotropy in obtained superconducting
parameters can be approximately estimated from the anisotropic GL relation:
\begin{equation}
\gamma _{anis}=\sqrt{\Gamma }=\frac{H_{c2}^{ab}}{H_{c2}^{c}}=\frac{\xi _{ab}%
}{\xi _{c}}=\frac{\lambda _{c}}{\lambda _{ab}}=\frac{\kappa _{ab}}{\kappa
_{c}}\sim\frac{H_{c1}^{c}}{H_{c1}^{ab}}
\end{equation}%
where $\Gamma =\frac{m_{c}^{\ast }}{m_{ab}^{\ast }}=\frac{\rho _{c}}{\rho
_{ab}}$ with the effective electron mass $m_{i}^{\ast }$ along the $i$
direction. The obtained mass tensor anisotropy $\Gamma \sim$ 7 is
comparable with the experimental resistivity ratio $\sim$ 10 of ZrTe$_{3}$.
\cite{21} This points to small anisotropy in the gap function $\Delta $(k$
_{F}$).\cite{37} We list the obtained superconducting parameters in Table
1.

\begin{table*}[tbp] \centering%
\caption{Superconducting State Parameters for Cu$_{0.05}$ZrTe$_{3}$.}
\begin{tabular}{cccccccc}
\hline\hline
Cu$_{0.05}$ZrTe$_{3}$ & &$H_{c2}^{i}(0)$ (kOe) & $H_{c1}^{i}(0)$ (Oe) & $%
H_{c}(0)$ (Oe) & $\kappa _{i}$(0) & $\xi _{i}$(0) (nm) & $\lambda _{i}$(0)
(nm) \\
\hline
\multirow{2}*{$i=ab$} &$onset$& 32(4) & \multirow{2}*{8.3(1)}    & 240(30) & 93(14) & 16(1.2) & 605(210)   \\
                      &$mid$  & 27(3) &                       & 225(27) & 85(11) & 18(1.8) & 570(184)   \\
\hline
\multirow{2}*{$i=c$} &$onset$ & 13(2) & \multirow{2}*{27.7(4)}    & 330(50) & 28(5) & 6.5(1.3)& 1490(420)  \\
                     &$mid$   & 10(2) &                        & 295(50) & 24(5) & 6.7(1.3)& 1539(477)  \\
 \hline\hline
\end{tabular}%
\label{TableKey}%
\end{table*}%

\begin{figure}
\includegraphics[width=0.45\textwidth,angle=0]{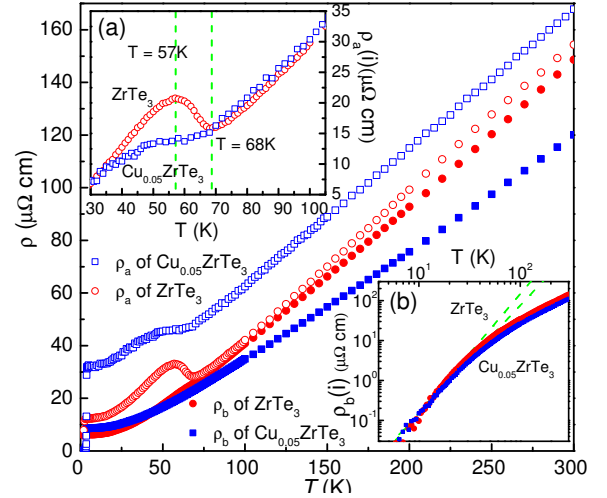}
\caption{Temperature dependence of $\rho_{a}$ and $\rho_{b}$ for ZrTe$_{3}$ and Cu$_{0.05}$ZrTe$_{3}$.
(a) Temperature dependence of $\rho_a$($i$) for ZrTe$_{3}$ and Cu$_{0.05}$ZrTe$_{3}$;
the peak (T $=$ 57 K) and the valley (T $=$ 68 K) of CDW anomaly in ZrTe$_3$ are marked with two dashed green lines.
(b) Logarithmic plots of temperature dependence for $\rho_b$($i$) with fitting results (dashed green lines) for the Umklapp scattering process (see text).}
\end{figure}

The in-plane anisotropic electronic transport properties in the normal state of Cu$_0.05$ZrTe$_3$ and ZrTe$_3$ are studied. Fig. 4 shows the temperature dependence of $\rho_a$ and $\rho_b$ for ZrTe$_3$ and Cu$_x$ZrTe$_3$, which indicates the absolute value of resistivity is quasi 2D for both of them. Since the residual resistivity ($\rho _0$) is related to the temperature independent defect scattering contribution, resistivity subtracted by $\rho _0$ ($\rho($i$)$ = $\rho$ - $\rho_0$) can provide information about CDW or scattering mechanism. Compared with ZrTe$_3$ (Fig. 4 (inset,a)), a weaker anomaly below 68 K in $\rho_{a}($i$)$ for Cu$_{0.05}$ZrTe$_{3}$ can be observed without a obvious shift. This is in contrast to shift of anomaly in ZrTe$_{3}$ under high pressure\cite{27} and 1T-TiSe$_2$ by Cu intercalation or under high pressure\cite{5,6}. Since the resistivity anomaly (shown in Fig. 4 (inset, (a)) comes from of the reduction of DOS at FS due to the CDW bandgap, the weaker anomaly reflects that the CDW bandgap is partially filled via Cu intercalation. Meanwhile, the Cu intercalation has no significant effect on $\rho_{b}($i$)$ (shown in Fig. 4 (inset, b)), even though the absolute values of resistivity are quasi 2D\cite{21}. The scattering along the $\widehat{b}$ - axis in the intercalated and the pure material are not related to the CDW. Linear relation of $\rho_b(i)$ shows that the dominant scattering channel along Zr-Zr metal chains in both materials is electron-electron Umklapp scattering (Fig. 4 (inset, b)), as expected for quasi one dimensional chains\cite{38}. The Umklapp process will result in a $\rho(T)$ = $\rho_0$ + AT$^{n}$ (where 2 $\leq $ n $\leq $ 3) for k$_{B}$T $ < $ (0.1 - 0.3)$\left\vert \delta \right\vert $, and a linear relation at high temperature for k$_{B}$T $>$ $\left\vert\delta\right\vert$\cite{38}. The parameter A is a constant and the $|\delta |$ is the typical inter-chain interaction energy. The power law temperature dependence is observed for T $<$ 25 K with $n = 2.98 \pm 0.03$ and $n = 2.70 \pm 0.03 $ for ZrTe$_{3}$ and Cu$_{0.05}$ZrTe$_{3}$, respectively (Fig. 4 (inset,b)). The estimated $|\delta|$\ range is (100 $-$ 250) K, which is consistent with the observed linear $\rho_b$(T) above 180 K (Fig.4).

Generally speaking, anisotropic lattice expansion via Cu intercalation is different from the isotropic high pressure in ZrTe$_3$. It reminds us that uniaxial pressure in NbSe$_3$ results in different behavior with isotropic high pressures\cite{39}, and that most CDW bearing materials are anisotropic low dimensional system. Thus, the isotropic high pressure provides a combined effect in these materials, rather than a simple way of band widening as expected. It appears that ZrTe$_3$ is an good candidate for the systematic study of the interplay and of CDW and SC, and further spectroscopic measurements would be of interest in order to unfold the details of the phonon mode softening and CDW melting as a function of Cu intercalation. Since SC is induced in another layered tritellurides TbTe$_3$ under high pressure\cite{40}, it is worthwhile to search SC by doping or intercalation in the family of CDW bearing layered rare earth tritellurides as well.

In summary, we have shown that Cu intercalation in ZrTe$_{3}$ results in the
expansion of the lattice parameters in the $\widehat{a}$ - $\widehat{c}$
direction (orthogonal to Zr - Zr metal chains) and partial filling of the CDW energy gap. Bulk SC with \textit{T}$_{c}$ = 3.8 K is discovered in Cu$_{0.05}$ZrTe$_{3}$, and its superconducting parameters are given. Along the Zr-Zr metal chains, electrical transport is not related to CDW, and the dominant dominant scattering mechanism is the electron - electron Umklapp scattering.

We thank John Warren for the help with SEM measurements. This work was
carried out at BNL, which is operated for the U.S. Department of Energy by
Brookhaven Science Associates DE-Ac02-98CH10886.


\begin{thebibliography}{99}
\bibitem{1} R. E. Peierls, Ann. Phys. Leipzig \textbf{4}, 121 (1930).

\bibitem{2} G. Gruner, Density Waves in Solids, Addison-Wesley, Reading MA 1994.

\bibitem{3} P. Monceau, et al., Phys. Rev. Lett. \textbf{39}, 161 (1977).

\bibitem{4} R. J. Cava, et al. Nature \textbf{332}, 814 (1988).

\bibitem{5} E. Morosan, et al., Nature Physics \textbf{2}, 544 (2006).

\bibitem{6} A. F. Kusmartseva, et al., Phys. Rev. Lett. \textbf{103}, 236401 (2009).

\bibitem{7} T. Kiss et al., Nature Physics \textbf{3}, 720, (2007).

\bibitem{8} Qimiao Si and Frank Steglich, Science \textbf{329}, 1161 (2010).

\bibitem{9} P. Gegenwart, et al., Nature Physics\textbf{4}, 186 (2008).

\bibitem{10} S. E. Sebastian, et al.,Proc. Nat. Ac. Sc. \textbf{107}, 6175 (2010).

\bibitem{11} P. Monthoux, et al., Nature \textbf{450}, 1177 (2007).

\bibitem{12} D. J. Scalapino, Science \textbf{284}, 1282 (1999).

\bibitem{13} H. Barath, et al., Phys. Rev. Lett. \textbf{100}, 106402 (2008).

\bibitem{14} K. Sun, et al.,Phys. Rev. B \textbf{78}, 085124 (2008).

\bibitem{15} M. Kim, et al., Advanced Mater. \textbf{22}, 1148 (2010).

\bibitem{16} J. F. Zhao, et al., Phys. Rev. Lett. \textbf{99}, 146401 (2007).

\bibitem{17} S. Y. Li, et al., Phys. Rev. Lett. \textbf{99}, 107001 (2007).

\bibitem{18} S. Hellmann, et al., Phys. Rev. Lett. \textbf{105}, 187401 (2010).

\bibitem{19} S. Furuseth, et al., Acta Chem. Scand., Ser. A \textbf{29}, 623 (1975).

\bibitem{20} K. St\"{o}we and F. R. Wagner, J. Solid State Chem. \textbf{138}, 160 (1998).

\bibitem{21} S. Takahashi, et al., Solid State Commun. \textbf{49}, 1031 (1984).

\bibitem{22} D. J. Eaglesham, et al., J. Phys. C: Solid State Phys. \textbf{17}, L697 (1984).

\bibitem{23} C. Felser, et al., J. Mater. Chem. \textbf{8}, 1787 (1998).

\bibitem{24} P. Starowicz, et al., J. Alloys and Compounds \textbf{442}, 268 (2007).

\bibitem{25} M. Hoesch, et al., Phys. Rew. B \textbf{80}, 075423 (2009).

\bibitem{26} H. Nakajima, et al., Physica B+C \textbf{143}, 240 (1986).

\bibitem{27} R. Yomo, et al., Phys. Rev. B \textbf{71}, 132508 (2005).

\bibitem{28} K. O. Klepp, et al., Z. Naturforsch., B: Chem. Sci. \textbf{57}, 1265 (2002).

\bibitem{29} W. Finckh, et al., J. Alloys and Compounds \textbf{262-263}, 97 (1997).

\bibitem{30} B. Hunter, Int. Un. of Cryst. Comm. Newsletter \textbf{20} (1998).

\bibitem{31} H. Kupfer, et al., Cryogenics \textbf{28}, 650 (1988).

\bibitem{32} C. Senatore, et al., Phys. Rev. B \textbf{78}, 054514 (2008).

\bibitem{33} W. L. McMillan, Phys. Rev. \textbf{167}, 331 (1968).

\bibitem{34} X. D. Zhu, et al., J. Phys.: Condens. Matter \textbf{21}, 145701 (2009).

\bibitem{35} N. R. Werthamer, et al.,Phys. Rev. \textbf{147}, 295 (1966).

\bibitem{36} J. R. Clem, Physica C \textbf{162-164}, 1137 (1989).

\bibitem{37} P. Miranovic, et al., J. Phys. Soc. Japan \textbf{72}, 221 (2003).

\bibitem{38} A. Oshiyama, et al., J. Phys. Soc. Jpn. \textbf{45}, 1136 (1978).

\bibitem{39} Kh. B. Chashka, et al., Physica B \textbf{203} 75 (1994).

\bibitem{40} J. J. Hamlin, et al., Phys. Rev. Lett. \textbf{102}, 177002 (2009).


\end{thebibliography}
\end{document}